\documentclass[]{aa520}
\usepackage{natbib,amssymb,graphicx,graphics,psfig}
\bibpunct{(}{)}{;}{a}{}{,} 
\begin{document}

\title{The mineral composition and spatial distribution of the dust
  ejecta of NGC~6302 \thanks{Based on observations with ISO, an ESA
    project with instruments funded by ESA Member States (especially
    the PI countries: France, Germany, the Netherlands and the United
    Kingdom) and with the participation of ISAS and NASA}
  \thanks{Based on observations made with ESO Telescopes at the La
    Silla or Paranal Observatories under programme ID   67.D-0132(A)}}

\author{F.~Kemper\inst{1}, F.J.~Molster\inst{2}, C.~J\"ager\inst{3},
  L.B.F.M.~Waters\inst{1,4}}

\institute{Astronomical Institute 'Anton Pannekoek', University of
  Amsterdam, Kruislaan 403, 1098 SJ Amsterdam, The Netherlands \and
  ESTEC/ESA, RSSD-ST, Keplerlaan 1, 2201 AZ Noordwijk, The Netherlands
  \and University of Jena, Astrophysical Institute and University
  Observatory (AIU), Schillerg\"asschen 3, D-07745 Jena, Germany \and
  Instituut voor Sterrenkunde, K.U. Leuven, Celestijnenlaan 200B, 3001
  Heverlee, Belgium}

\offprints{F.~Kemper ({\tt ciska@science.uva.nl})}

\date{received\dots, accepted\dots}

\authorrunning{F. Kemper et al.}  \titlerunning{The mineralogy of
  NGC~6302}

\abstract{We have analysed the full ISO spectrum of the planetary
  nebula \object{NGC~6302} in order to derive the mineralogical
  composition of the dust in the nebula. We use an optically thin dust
  model in combination with laboratory measurements of cosmic dust
  analogues.  We find two main temperature components at about 100 and
  50 K respectively, with distinctly different dust compositions. The
  warm component contains an important contribution from dust without
  strong infrared resonances. In particular the presence of small warm
  amorphous silicate grains can be excluded.  The detection of weak
  PAH bands also points to a peculiar chemical composition of the dust
  in this oxygen-rich nebula. The cool dust component contains the
  bulk of the mass and shows strong emission from crystalline
  silicates, which contain about 10 percent of the mass. In addition,
  we identify the 92 $\mu$m band with the mineral calcite, and argue
  that the 60 $\mu$m band contains a contribution from the carbonate
  dolomite. We present the mass absorption coefficients of six
  different carbonate minerals. The geometry of the dust shell around
  \object{NGC~6302} is studied with mid-infrared images obtained with
  TIMMI2. We argue that the cool dust component is present in a
  circumstellar dust torus, while the diffuse emission from the warm
  component originates from the lobes.  }

\maketitle

\section{Introduction}
\label{sec:intro}

All low and intermediate mass stars end their life on the Asymptotic
Giant Branch (AGB) by ejecting their entire H-rich envelope through a
slowly expanding, dusty wind. After the AGB, the central star quickly
increases its effective temperature to values high enough to begin
ionizing its AGB ejecta: a planetary nebula (PN) is born. Depending on
the mass and thus luminosity of the star, the transition from AGB to
PN can go very fast: for the most massive objects, in less than 1000
years ionization of the nebula begins. These massive objects therefore
are characterized by dense AGB remnants and a very hot luminous
central star.

\object{NGC~6302} is probably one of the best studied PNe with a
massive progenitor. A recent determination of the mass of the ionized
nebula is about 2~$M_{\odot}$ \citep{PB_99_NGC6302}, based on a
distance determination of 1.6 kpc \citep{GRM_93_ngc6302}. Dust mass
estimates also indicate a very massive shell ($M_{\mathrm{d}} = 0.02
\, M_{\odot}$), using a distance of 2.2 kpc \citep{GMR_89_NGC6302}.
These distance determinations rely on VLA observations of the
expansion of the nebula over a 2.75 yr period, assuming an outflow
velocity of 13 km s$^{-1}$. In principle, this is a very reliable
method, however the epoch over which the nebula is observed, is very
short.  Therefore the increase in size is very small and hard to
measure, indicating that the error bars on these distance
determinations are still very large.  Instead, we adopt a distance of
0.91 kpc, based on emission-line photometry of \object{NGC~6302}, from
which the $B$ and $V$ magnitudes, the luminosity and the distance can
be derived \citep{SK_89_PNe}. The large number of luminous PNe studied
gives confidence in the distance determination from this statistical
method.

The nebular abundances indicate that \object{NGC~6302} is a type I
nebula, consistent with a massive progenitor.
\citet{CRB_00_ngc6302_ngc6537} estimate that the progenitor mass is
4--5 $M_{\odot}$.  The morphology of the nebula observed at optical
wavelengths is highly bipolar, pointing to non-spherical mass loss on
the AGB, resulting in a dusty torus in the equatorial region
\citep{LD_84_ngc6302}. The inclination angle of the system is $\sim
45^{\mathrm{o}}$ with respect to the line-of-sight
\citep{B_94_ngc6302}. The temperature of the central star is very
high: \citet{CRB_00_ngc6302_ngc6537} mention a temperature of $\sim
250\,000$ K, while \citet{PBD_96_centralstar} arrive a temperature of
$\sim 380\,000$ K. Although there is some uncertainty about the
distance and therefore masses and luminosities involved, everything
points to a rather massive and luminous progenitor.

The Infrared Space Observatory (ISO) spectrum has been presented in
several papers \citep{B_98_LWS_AGB,MLS_01_NGC6302,MWT_02_xsilI} and is
characterized by a wealth of narrow solid state features in the 20--70
$\mu$m spectral range caused by circumstellar dust in the AGB remnant,
as well as by strong emission lines from a multitude of fine-structure
lines originating from the ionized gas in the nebula. The dust bands
have been identified with crystalline silicates and a number of other
components \citep{KTS_00_diopside,MLS_01_NGC6302,MWT_02_xsilII}.
\citet[hereafter referred to as Paper~I]{KJW_02_carbonates} have
reported on the detection of carbonates in the dust shell, based on
the identification of broad features at $\sim$60 and $\sim$92 $\mu$m.

Optical and near-infrared images have already shown that the dust
distribution around \object{NGC~6302} is rather complex
\citep{LD_84_ngc6302,SCM_92_PNimages}.  The ISO spectroscopy supports
this, because a broad dust temperature range is required to explain
the shape of the spectral energy distribution. In addition the dust
composition seems to be very complex, as there is evidence for a mixed
chemistry by the presence of both oxygen-rich dust and carbon-rich
dust features, the latter in the form Polycyclic Aromatic Hydrocarbons
(PAHs) (see \citet{MLS_01_NGC6302} for a discussion on the origin of
this dichotomy).

In order to reconstruct the mass loss history of \object{NGC~6302},
including the geometry and composition of the AGB wind, infrared
spectroscopy and imaging are needed.  Unfortunately, the ISO data lack
spatial information, limiting the analysis to the bulk dust
composition.  Mid-infrared imaging can reveal the present-day geometry
of the dust envelope, which puts limits on the mass loss history.

This paper is organized as follows: In Sect.~\ref{sec:TIMMI2}
the ground-based mid-infrared imaging and spectroscopy is presented.
The ISO spectroscopy is discussed in Sect.~\ref{sec:ISO}, which describes
our laboratory data of carbonates (Sect.~\ref{sec:carbonates}) and a
model fit to the observed spectrum (Sect.~\ref{sec:modelfit}).
In Sect.~\ref{sec:discussion} we propose a possible geometry of
\object{NGC~6302} and discuss the astronomical relevance of carbonates.
Our results are summarized in Sect.~\ref{sec:summary}.

\section{The TIMMI2 observations}
\label{sec:TIMMI2}

\subsection{N- and Q-band imaging}
\label{sec:nqimag}

We have observed \object{NGC~6302} using the TIMMI2 mid-infrared
imaging spectrograph attached to the 3.6m telescope at the European
Southern Observatory, La Silla, Chile. For a description of the
instrument see \citet{RLW_00_TIMMI2}. The observations were carried
out on the night of June 17/18, 2001. Observing conditions were not
optimal, with occasional clouds and variable seeing (between about
0.6$''$ and 1.4$''$).  Imaging was done in four photometric bands,
centred at 9.8 (N2), 10.4, 11.9, and 20 (Q) $\mu$m.  In addition, we
obtained two 8.35-11.55 $\mu$m long-slit spectra. We observed
\object{HD~81797} and \object{HD~169916} as point sources in order to
determine the shape of the point spread function, as well as for flux
calibration purposes, which is only performed in case of spectroscopy.

To correct for background, chopping and nodding has been performed.
Apart from the Q-band image, the chopping and nodding images fell all
inside the frame. For the various N-band images we used a lens scale
corresponding to a pixel size of 0.3$''$/pixel, whereas the Q-band
image was obtained with a pixel size of 0.2$''$/pixel.  The dimensions
of the detector are 320$\times$240 pixels of which the central
300$\times$220 pixels were used for further analysis. The chop throw
used for the images in the N-band is 30$''$, with a nodding offset of
30$''$ perpendicular on the chop direction. For the Q-band image, we
applied a chop throw of 40$''$, and a nodding offset of 40$''$ in the
same direction as the chop.

Both the positive and negative images were used for the resulting
image. The final image has been sharpened by means of a deconvolution
using the point spread function (PSF) derived from \object{HD~81797}
for the Q-band.  Unfortunately, for the N2, 10.4 and 11.9 filters no
standard star was measured in the same night with the same settings,
and we used observations of standards of another night. For the N2 and
10.4 filters no standards were available at all. Therefore, we used
\object{HD~169916} measured with the 11.9 $\mu$m filter for all three
filters for the PSF, assuming that the PSF would be similar in all
three N-bands. A comparison with other N2-band images (with different
settings, but the positive and negative images did overlap each other)
showed that our assumption is not unreasonable.

\begin{figure*}
  \sidecaption \includegraphics[width=12cm]{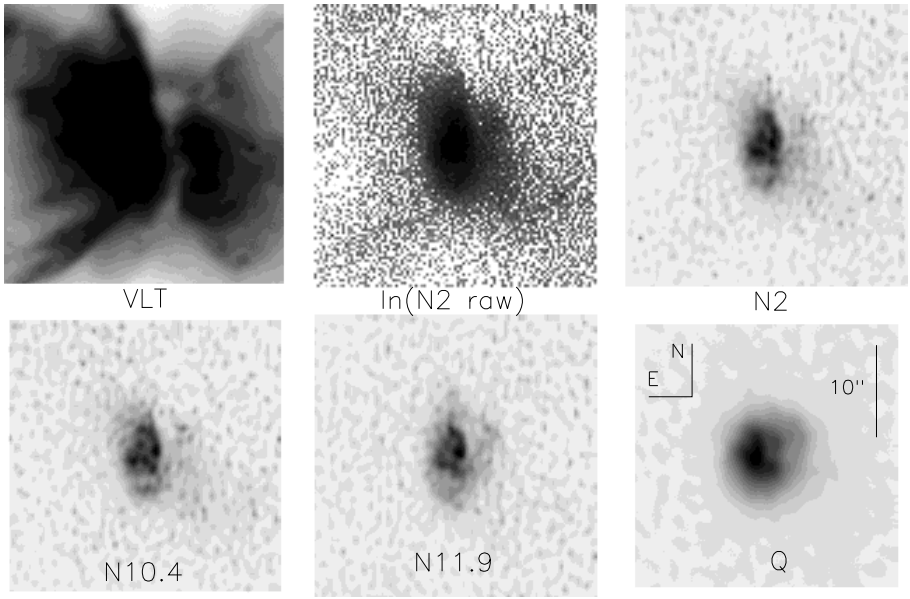}
\caption[]{TIMMI2 images of \object{NGC~6302}. In the upper left corner
  the VLT-first light image of \object{NGC~6302} is shown
  \citep{ESO_98_VLTfirstlight}. The upper middle image is the
  logarithmic non-deconvolved N2 band image, which shows the low
  brightness structure somewhat better than the deconvolved images.
  In the upper right corner the deconvolved N2 band image is shown, in
  which detailed structure in the centre appears.  The bottom row
  shows the deconvolved 10.4, 11.9 and Q-band images. All images have
  the same orientation and scale.}
\label{fig:TIMMI2}
\end{figure*}

\begin{figure}
\resizebox{\hsize}{!}{\includegraphics{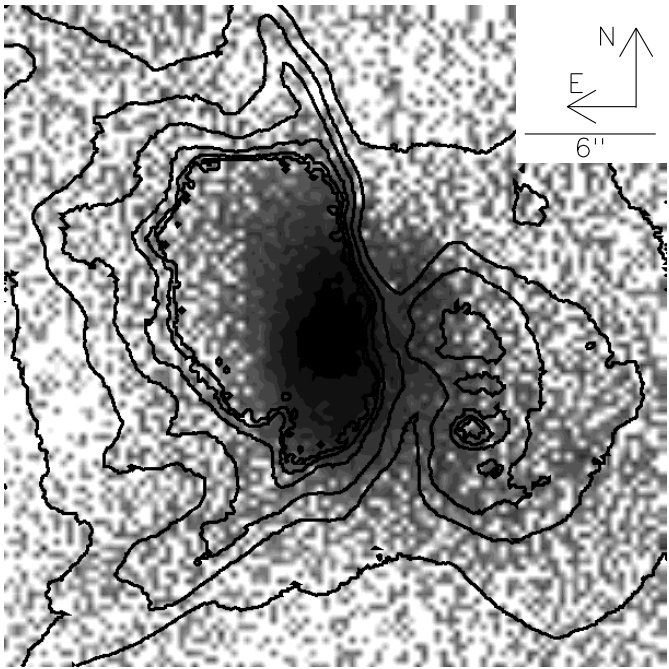}}
\caption[]{TIMMI2 N2-band image of \object{NGC~6302}, overlaid with the
  contours of the optical VLT image (Fig.~\ref{fig:TIMMI2}).}
\label{fig:overlay}
\end{figure}

Fig.~\ref{fig:TIMMI2} shows the final TIMMI2 images at N2, N10.4,
N11.9, and Q band, as well as the "raw" (i.e. before
deconvolution) N2 image, and an optical VLT image of the nebula
for comparison \citep{ESO_98_VLTfirstlight}. The
"raw" N2 image shows a bright elongated region of approximately 10$''$
by 4$''$, and a much fainter region offset to the south and
west. There is also evidence for faint arcs of emission that are
reminiscent of the X-shaped nebula seen in the optical image. In
fact, the mid-IR emission has a morphology which is strikingly
similar to that seen in the optical (see Fig.~\ref{fig:overlay}).
This is surprising since the mid-IR emission originates from
(warm) dust, while the optical emission is dominated by ionised
gas. The equatorial torus, seen as a waist in the optical
images, must contain very cold dust which does not yet contribute
to the flux at 10 $\mu$m. We will return to this point when we
discuss the ISO spectrum.

The deconvolved N2, N10.4 and 11.9 $\mu$m images in
Fig.~\ref{fig:TIMMI2} are very similar and show a rather "blobby"
appearance. Note that the faint structure in the background is a
common artifact of image deconvolution. However, the structure in the
central area is real, since we see similar substructure in different
observations deconvolved with different point spread functions. Also
the 6cm radio map of NGC 6302 shows substructure in this region
\citep{GMR_89_NGC6302}. Our images are very similar to the 3.3 $\mu$m
image taken by \citet{CRB_00_ngc6302_ngc6537}, taking into account
that the orientation of their image is accidentally rotated by
180$^{\rm o}$ (Casassus, priv.~comm.).

The deconvolved Q band image (Fig.~\ref{fig:TIMMI2}) tends to show
more spherical emission, although the eastern part of the nebula is
still dominating the emission. Due to the reduced spatial resolution
at 20 $\mu$m it is not possible to say if the emission is equally
blobby as seen at 10 $\mu$m.

\subsection{N-band spectroscopy}
\label{sec:nspec}

\begin{figure}
\resizebox{\hsize}{!}{\includegraphics{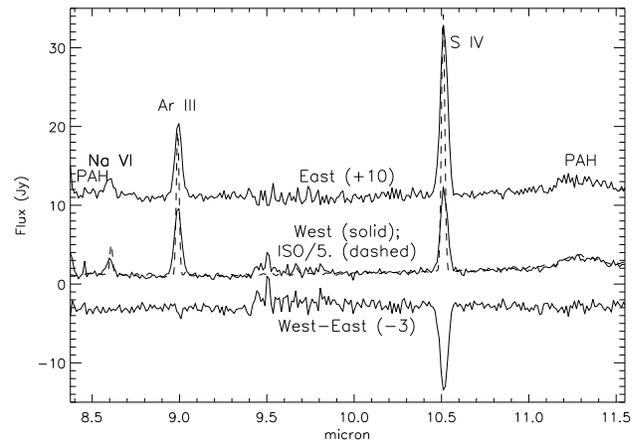}}
\caption[]{TIMMI2 N-band spectra \object{NGC~6302}. The spectra were taken
  1.2$''$ east and 1.2$''$ west of the brightest (east of the torus) lobe
  in the image. For comparison the ISO spectrum (divided by 5) is
  given as a dashed line. The relative flux calibration is quite
  accurate, but the absolute flux calibration is only certain up to
  40\%. The difference spectrum of the west and east side of the lobe
  is also given. The lines around 9.5 $\mu$m are due to telluric
  lines.}
\label{fig:TIMMI2spec}
\end{figure}

\begin{figure}
\resizebox{\hsize}{!}{\includegraphics{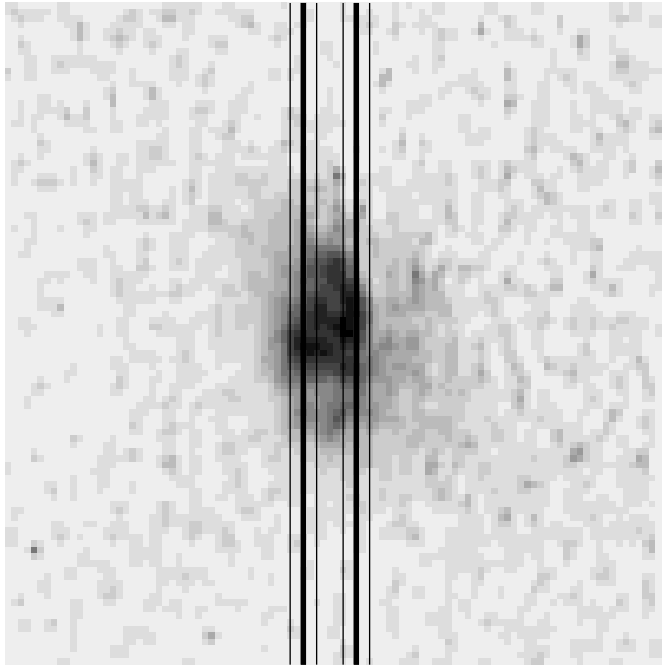}}
\caption[]{The position of the N-band slit showed on the deconvolved N2 image.
  North is up and east is to the left. Both slits were
  $1.2''\times70''$ and placed 1.2$''$ east and west from the central
  peak. It is clear that both slit positions cover the eastern lobe of
  the nebula.}
\label{fig:slitpos}
\end{figure}

In Fig.~\ref{fig:TIMMI2spec} we show the spectra taken at the slit
positions indicated in Fig.~\ref{fig:slitpos}. We used
\object{HD~169916} as a reference spectrum, assuming that this
star has a $F(\nu) \sim \nu^2$ flux distribution. The comparison
with other calibration observations showed that the absolute flux
calibration is only accurate up to about 40\%. The relative flux
calibration is much more accurate. The wavelength calibration
provided by ESO on the TIMMI2 home page appeared to be inaccurate,
and we determined a new wavelength calibration, based on the
[Ar{\sc iii}] and [S{\sc iv}] line, assuming a linear pixel to
wavelength dependence. The fact that the [Na{\sc vi}] line and the
PAH features are now at the right wavelengths shows that our
assumption is valid (using the calibration files provided by ESO
the 11.3 $\mu$m PAH feature was found at 11.7 $\mu$m). The
comparison with the ISO spectrum (dashed line in
Fig.~\ref{fig:TIMMI2spec}), gives confidence in the new wavelength
calibration.

The spectrum is dominated by the forbidden emission lines of [S{\sc iv}]
and [Ar{\sc iii}].  However the PAH features at 11.3 and also 8.6 $\mu$m
are also found. The noise around 9.5 $\mu$m is due to telluric lines.

The east and west slit are both probing the eastern lobe of the nebula
(Fig.~\ref{fig:slitpos}). The difference spectrum is very flat, apart
from the [S{\sc iv}] line. The fact that even around 8.6 and 11.3 $\mu$m
it is flat, indicates that the distribution of the PAHs is quite
constant in the bright eastern lobe. This is in agreement with the 
3.3 $\mu$m image obtained by \citet{CRB_00_ngc6302_ngc6537}.

\section{Analysis of the ISO spectrum}
\label{sec:ISO}

\begin{figure}
  \resizebox{\hsize}{!}{\includegraphics{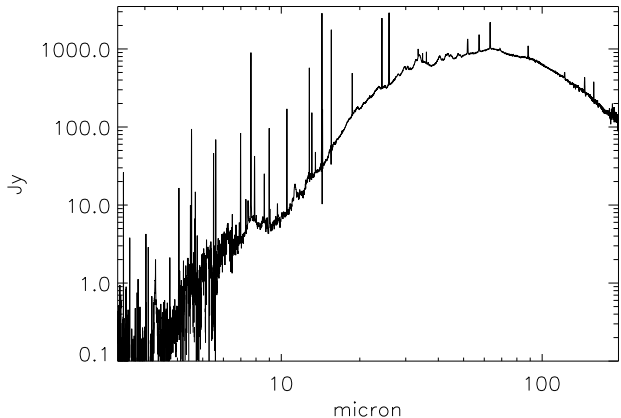}}
\caption[]{The ISO 2.4 -- 200 $\mu$m spectrum of \object{NGC~6302}.
This figure is
  based on data presented in \citet{MLS_01_NGC6302}. Clearly visible
  are the numerous emission lines due to fine-structure transitions in
  the ionized gas.}
\label{fig:iso}
\end{figure}

The ISO SWS and LWS spectra have been presented by
\citet{B_98_LWS_AGB} and \citet{MLS_01_NGC6302,MWT_02_xsilI} and in
Paper~I. We briefly repeat the overall characteristics of the
spectrum. The SWS and LWS apertures, which have sizes in the range
14$\times$20$''$ to 80$''$, cover a large fraction of the nebula and
therefore provide no detailed spatial information. The spectrum, shown
in Fig.~\ref{fig:iso}, is dominated by a strongly rising spectrum
which peaks near 55 $\mu$m, and numerous, sometimes rather narrow
emission bands. Both the continuum and these emission bands are caused
by the circumstellar dust. In addition, strong, spectrally unresolved
emission from fine-structure lines are evident, which are caused by
the ionized nebular gas. These lines can be used to determine the
chemical abundance of the gas and properties of the central star
\citep[e.g.][]{PBD_96_centralstar,PB_99_NGC6302}.  In this work we
focus on the analysis of the dust component.

\subsection{Laboratory spectroscopy of carbonates}
\label{sec:carbonates}

\begin{table}
\caption[]{Chemical formula, crystal system and content of impurities of
the carbonate minerals determined by SEM and EDX measurements. The
impurities in dolomite, magnesite, siderite and ankerite are the carbonates
corresponding to the metals mentioned in the last column.
The used abbreviations in column 3
describe the crystal system of the mineral (trigonal and
rhombic).}
\vspace*{0.2cm}
\label{tab:chemcomp}
\begin{tabular}{lccl}
\hline
\hline
Mineral     & Formula          & Crystal    & Impurities \\
            &                  & system     &   (wt\%)\\
\hline
dolomite    & CaMg(CO$_3$)$_2$ &   tri.     & $\sim$0.5 Fe, very pure \\
calcite     & CaCO$_3$         &   tri.     &  $<$0.1 Si, very pure  \\
aragonite   & CaCO$_3$         &   rhom.    & $<$0.2 Si, very pure  \\
magnesite   & MgCO$_3$         &   tri.     & $\sim$ 1.0 Fe; 0.2-0.9 Ca  \\
siderite    & FeCO$_3$         &   tri.     & $\sim$ 2.0 Mg; $\sim$ 5.0 Mn \\
ankerite    & CaFe(CO$_3$)$_3$ &   tri.     & 2-3 Mn; 3.5-4.2 Mg \\
\hline
\hline
\end{tabular}
\end{table}

\citet{MLS_01_NGC6302} provide a useful inventory of the solid state
emission bands in the ISO spectrum of \object{NGC~6302}, and we will
not repeat their analysis here. While many emission bands were
successfully allocated to (mostly oxygen-rich) dust components,
several prominent emission bands, notably near 29, 48, 60 and 90
$\mu$m remained partially unidentified. We have investigated possible
carriers for the $\sim$92 $\mu$m band, reported to be present in
\object{NGC~6302} and \object{NGC~6537}
\citep{MLS_01_NGC6302,MWT_02_xsilI}. During the course of our
laboratory studies of cosmic dust analogues, we noted that the
carbonate calcite (CaCO$_3$) has a strong resonance at $\sim$92
$\mu$m. This prompted us to carry out a systematic laboratory study of
transmission spectra of several carbonate species. A series of
carbonates of possible cosmic relevance has been chosen for laboratory
investigations, including the natural minerals ankerite (origin
unknown), aragonite (Bohemia), dolomite (Eugui, Navarra, Spain),
calcite (origin unknown), magnesite (Steiermark, Austria) and siderite
(Sch\"onbrunn, Vogtland, Germany). The purity and homogeneity of the
natural samples are investigated by scanning electron microscopy (SEM)
and energy dispersive X-ray analysis (EDX). The macroscopic phase
homogeneity of the carbonate samples was measured by X-ray
diffraction. The chemical composition and the content of impurities
are presented in Table~\ref{tab:chemcomp}.

The minerals dolomite, calcite, aragonite and magnesite turned out to
be very pure and homogeneous. The impurities in siderite and ankerite
are the corresponding carbonates mentioned in Table~\ref{tab:chemcomp}.
These are homogeneously distributed within the main mineral component.

\begin{figure}
\resizebox{\hsize}{!}{\includegraphics{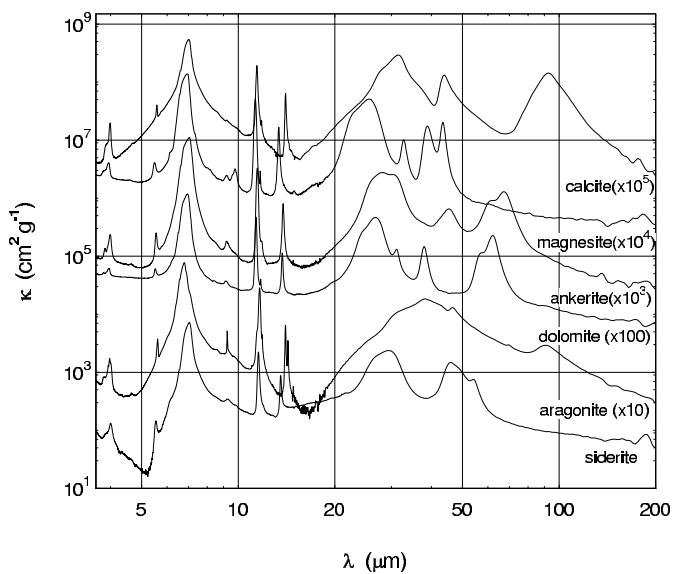}}
\caption[]{Mass absorption coefficients of natural carbonates
  derived from transmission measurements of the powders embedded in
  KBr and PE. For clarity, the spectra have been multiplied by the
  factor indicated in the plot.}
\label{fig:IRtrans}
\end{figure}

Small carbonate grains with an average size smaller than 2 $\mu$m have
been produced by grinding and sedimentation in water-free acetone. In
order to measure the infrared spectrum over a wide spectral range the
small grains were embedded in transparent material like KBr and
Polyethylene (PE). Infrared spectroscopy was performed using a Bruker
113v Fourier transform spectrometer in the wavelength range between
2-200 $\mu$m. The infrared spectral behaviour of the carbonate
minerals is demonstrated in Fig.~\ref{fig:IRtrans}, and can be found
in digitized form in the Jena-St.~Petersburg Database of Optical
Constants \citep{HIK_99_database}.

Magnesite, siderite, ankerite, dolomite, and calcite belong to the
carbonates of the calcite group and crystallize in the trigonal
crystal system. Carbonates containing cations with larger radii than
Ca$^{2+}$ crystallize in the rhombic crystal system. CaCO$_3$ is
dimorph and can exist in both structures as calcite and aragonite. The
typical crystalline structure of the carbonates in the calcite group
is characterized by an alternating sequence of cation and anion
layers. In both magnesite and calcite, the plane carbonate group
consists of three oxygen located at the corners of an equilateral
triangle and a carbon in the centre. The bonding between C and O is
covalent whereas the bonding between the cations and the carbonate
anion (CO$_3$$^{2-}$) is mainly ionic.  Each cation is octahedrally
coordinated by six O atoms of the adjacent CO$_3$$^{2-}$ ion. The
crystals of the calcite group with mixed metal composition -- like
dolomite and ankerite -- show an alternating arrangement of the
calcium and magnesium or calcium and iron layers. In contrast to the
calcite structure the Ca$^{2+}$ ion in aragonite is coordinated by 9
O-atoms.

The infrared optical properties are determined by the crystalline
structure of the minerals. The spectral behaviour of carbonates has
been measured several times in previous studies
\citep[e.g.][]{HLP_70_carbonate,P_76_nonsilicates,OBF_97_carbonates,YSK_01_carbonates}.
The measured mid-infrared bands arise mainly from the carbonate anion.
The planar structure of this anion and its C$_3$ symmetry gives rise
to three infrared active modes at about 7, 11 and 14 $\mu$m. The 7
$\mu$m band can be attributed to the C=O stretching mode whereas the
11 and 14 $\mu$m bands are caused by the out-of-plane and in-plane
bending modes. The low frequency bands beyond 25 $\mu$m are due to
translation of the metal cations and represent motions perpendicular
and parallel to the plane of the carbonate anions
\citep{HLP_70_carbonate}.

\subsection{Dust model fit}
\label{sec:modelfit}

\subsubsection{Description of the model}
\label{sec:modeldescr}

We use the identified solid state components to quantitatively fit the
entire ISO spectrum. Inspection of the ISO data shows that all dust
bands are in emission, suggesting that the dust shell is optically
thin at infrared wavelengths. This substantially simplifies the
analysis. Given the complex geometry of the nebula, the multiple dust
components present in the spectrum, and the still incomplete knowledge
of the dust species present, we decided to focus our analysis on
determining the relative mass contributions of the known dust species,
rather than attempting a full 2-dimensional radiative transfer
calculation.  We use the optically thin nature of the dust at mid- and
far-infrared wavelengths to constrain the dust mass of the components
using mass absorption coefficients.  The thus derived dust masses will
not depend on the assumed geometry as long as the emission is
optically thin. The problem then reduces to deriving the temperature
over mass distribution of the various dust components
\citep[e.g.][]{BDV_00_ABAur}.  The advantage of using mass absorption
coefficients in an optically thin dust shell, is that it does not
require knowledge of the location of the dust.

We derive the dust mass and temperature distribution under the
following assumptions: {\bf i)} the dust shell is optically thin at
mid- and far-infrared wavelengths; {\bf ii)} all grains have the same
size, i.e.~0.1 $\mu$m; {\bf iii)} all grains are spherical; {\bf iv)}
all grains are of homogeneous composition; {\bf v)} the number density
of dust grains as a function of distance is a simple power law (see
Eq.~(\ref{eq:profiles}), where $\rho_0 = \frac{4}{3} \pi a^2
\rho_{\mathrm{d}} n_0$); {\bf vi)} the dust species all have the same
temperature ranges, limited by the temperatures at the inner radius
$T_0$ and at the outer radius $T_{\mathrm{max}}$, which are free
parameters. The result of our analysis is a mass over temperature
distribution $M_{\mathrm{d}}(T)$ of all identified dust species.  In
Sect.~\ref{sec:assumptions} we will discuss the validity and the
impact of these assumptions.

In the general case of an optically thin dust cloud, which
specifically applies to an optically thin dust shell, the specific
intensity emitted by the dust in the cloud can be written as

\begin{equation}
I_{\nu} (T) = \int_{V} \alpha_{\nu} \, B_{\nu}(T) \, dV \quad
\textrm{for} \quad \tau \ll 1
\end{equation}

where $\alpha_{\nu}$ is the specific absorptivity. By integrating over
volume $V$ this can be written as

\begin{equation}
I_{\nu} (T) = N \sigma_{\nu} B_{\nu}(T) \quad \textrm{or} \quad
I_{\nu} (T) = M_{\mathrm{d}}(T) \kappa_{\nu} B_{\nu}(T)
\end{equation}

with the absorption cross-section can be described in terms of the
geometric cross-section and absorption efficiency, according to
$\sigma_{\nu} = \sigma_{\mathrm{geom}} Q_{\nu}$. Following for
instance \citet{BDV_00_ABAur}, $M_{\mathrm{d}}(T)$ is the dust mass at
temperature $T$. The mass absorption coefficient is given by
$\kappa_{\nu}$ and the number of grains in the volume is represented
by $N$.

When we realize that we can write for the specific flux $F_{\nu} =
I_{\nu}/{D^2}$, and under the assumption that the dust grains are
identical spherical grains with a radius $a$, we find

\begin{equation}
F_{\nu} (T) = \frac{1}{D^2} \frac{1}{\rho_{\mathrm{d}}} \frac{3}{4}
\frac{1}{a} M_{\mathrm{d}}(T) Q_{\nu} B_{\nu}(T)
\label{eq:FNU}
\end{equation}

with $D$ the distance to the dust cloud and $\rho_{\mathrm{d}}$ the
density of the material in the dust grain.

In case of a circumstellar dust shell, we may assume that the
temperature and density distribution follow a power law:

\begin{equation}
T(r) = T_0 \Big( \frac{r}{r_0} \Big)^{-q}\\
\rho(r) = \rho_0 \Big( \frac{r}{r_0} \Big)^{-p}
\label{eq:profiles}
\end{equation}

where $r$ represents the distance to the central star, and $r_0$ the
inner radius of the dust shell. Using $dM_{\mathrm{d}} = 4 \pi r^2
\rho(r) dr$, we can express $M_{\mathrm{d}}$ in terms of the
temperature distribution in the dust shell, according to

\begin{equation}
M_{\mathrm{d}} (T(r))
= \frac{4 \pi \rho_0 {r_0}^3}{3-p}
\bigg( \frac{T(r)}{T_0} \bigg)^{-\frac{3-p}{q}}
\label{eq:MTr}
\end{equation}

For spherical dust grains the density at the inner radius is $\rho_0 =
\frac{4}{3} \pi a^3 \rho_{\mathrm{d}} n_0$, with $n_0$ the dust grain
number density at the inner radius, which is different for each
identified dust component.  The density $\rho_{\mathrm{d}}$
corresponds to the density of the grain material itself.  This can be
combined with Eqs.~(\ref{eq:MTr}) and~(\ref{eq:FNU}) to obtain an
expression of the specific flux:

\begin{equation}
F_\nu (T) = \frac{1}{D^2} \frac{4 \pi a^2 {r_0}^3 n_0}{3-p}
\bigg( \frac{T(r)}{T_0} \bigg)^{-\frac{3-p}{q}} Q_\nu B_\nu(T)
\label{eq:spec}
\end{equation}

where the observable total specific flux $\mathcal{F}_\nu$ is given by

\begin{equation}
\mathcal{F}_\nu = \int_{T_{0}}^{T_\mathrm{max}} F_\nu (T) \, dT
\label{eq:intspec}
\end{equation}

which can be compared directly to the ISO spectroscopy.

\subsubsection{Fit parameters and dust composition}
\label{sec:fitparameters}

Following the analysis of \citet{MLS_01_NGC6302}, we assumed that
enstatite (MgSiO$_3$), forsterite (Mg$_2$SiO$_4$) and crystalline
water ice are present, in addition to amorphous silicates. Diopside
((Ca,Mg)SiO$_3$) is believed to contribute to the 60 $\mu$m complex
\citep{KTS_00_diopside}.  Diopside is in the form of a solid solution
between wollastonite (CaSiO$_3$) and enstatite (MgSiO$_3$) in the
number ratio 46:54, i.e.~Wo46En54.  In addition, we considered
metallic iron as a dust component, which is found to be present in
the dust shells of evolved stars \citep{KDW_02_composition}, and the
carbonate species presented in this work.  Under the assumption that
all dust grains are spherical and have a radius $a$ of 0.1 $\mu$m, we
have derived $Q_\nu$ values from laboratory measurements of amorphous
olivine \citep{DBH_95_glasses}; metallic iron
\citep{HS_96_opacities}; forsterite and clino-enstatite
\citep{KTS_99_xsils}; diopside \citep{KTS_00_diopside}; water ice
\citep{BLW_69_ice,W_84_ice} and carbonates (this work).

\begin{figure}
  \resizebox{\hsize}{!}{\includegraphics{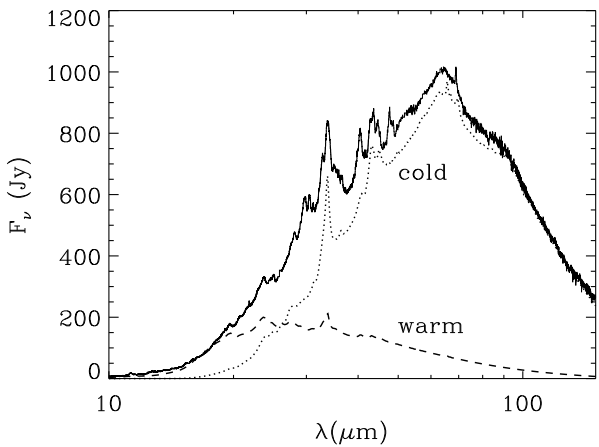}}
\caption[]{The warm and cool dust components required to fit the ISO
  spectrum of \object{NGC 6302}. The solid line represents the ISO
  spectrum.  The warm (100--118 K) and cool (30--60 K) dust components
  are represented by the labelled dotted and dashed lines.}
\label{fig:warmcool}
\end{figure}

\begin{table*}
\centering
\caption[]{Empirically derived dust masses by fitting
the ISO spectrum of \object{NGC~6302} for the temperature regimes 30--60 K and
100--118 K. The dust masses are obtained by integrating Eq.~(\ref{eq:MTr}) over
the considered temperature range.
The absolute accuracy of the dust masses is a factor 
of three, this is due to the uncertainties in laboratory measurements 
and inaccuracies in the used assumptions. The relative dust masses, 
i.e.~the mass ratios between the various identified species, are estimated 
to have an error of $\sim$50\% (see Sect.~\ref{sec:assumptions}). 
The values for the crystalline silicate components are lower limits, 
because we used laboratory measurements that were intrinsically much 
stronger than similar measurements obtained by other groups (Sect.~\ref{sec:assumptions}). 
In cases where lower limits are given, we did not detect the species 
indicated. The value indicates the
maximum amount that could be present without showing the spectral
features at a discernible level.}
\begin{tabular}{l|c c c|c c}
\hline
\hline
species            & $T$ (K)  & $ M_{\mathrm{d}}/M_{\odot}$        & mass fraction & $T$ (K)
 & $ M_{\mathrm{d}}/M_{\odot}$\\
\hline
amorphous olivine  & 60--30   & $4.7 \cdot 10^{-2}$   & 94 \%         & 118--100
 & $6.1 \cdot 10^{-6}$ \\
iron (or carbon)   &          &                       &               & 118--100
 & $1.2 \cdot 10^{-4}$ \\
forsterite         & 60--30   & $> 2.0 \cdot 10^{-3}$ & $>$ 4.0 \%    & 118--100
 & $3.7 \cdot 10^{-7}$ \\
clino-enstatite    & 60--30   & $> 5.5 \cdot 10^{-4}$ & $>$ 1.1 \%    & 118--100
 & $3.1 \cdot 10^{-7}$ \\
water ice          & 60--30   & $3.6 \cdot 10^{-4}$   & 0.72 \%       & 118--100
 & $< 1.5 \cdot 10^{-8}$\\
diopside           & 60--30   & $2.8 \cdot 10^{-4}$   & 0.56 \%       & 118--100
 & $< 1.2 \cdot 10^{-7}$\\
calcite            & 60--30   & $1.3 \cdot 10^{-4}$   & 0.26 \%       & 118--100
 & $< 1.0 \cdot 10^{-7}$\\
dolomite           & 60--30   & $7.9 \cdot 10^{-5}$   & 0.16 \%       & 118--100
 & $< 3.0 \cdot 10^{-8}$\\
\hline
\hline
\end{tabular}
\label{tab:dustmasses}
\end{table*}

Using Eqs.~(\ref{eq:spec}) and~(\ref{eq:intspec}) the far-infrared
spectrum has been fitted, under the assumption that all dust species
are found in the same temperature regimes. Thus, the free parameters
to be fitted are the number density at the inner radius $n_0$, which
differs for each dust component and scales with the mass contained in
that component; the temperature range defined by $T_0$ and
$T_{\mathrm{max}}$; and the slopes $p$ and $q$ of the density and
temperature power laws. For the distance $D$ we have used the measured
value of 0.91 kpc \citep{SK_89_PNe} and the grain size $a$ is assumed
to be 0.1 $\mu$m. The inner radius $r_0$ follows from the values of
$T_0$ and $q$, and is therefore not a fit parameter. If $p$ and $q$
are both chosen to be $-1/2$ the best results are obtained.
Inspection of the spectrum shows that the temperature distribution of
the dust must be very wide, which is incompatible with a simple
$r^{-2}$ density distribution of the dust.  We attempted to fit the
spectrum with a single range of temperatures but were not successful;
apart from a cool dust component (30--60 K), peaking near 55 $\mu$m,
we had to introduce a second dust component at 100--118 K dominating
the spectrum between 10 and 30 $\mu$m (see Fig.~\ref{fig:warmcool}).
The results of the fits are shown in Table~\ref{tab:dustmasses}, where
the mass of each dust component for both temperature ranges is given.
These masses are constrained by $n_0$ which is separately fitted for
all identified dust species and both temperature ranges, and are
obtained by integrating Eq.~(\ref{eq:MTr}) over the considered
temperature range.

The warm dust component could not consist of amorphous olivine,
because the strong resonances due to the Si-O stretching and O-Si-O
bending modes at 9.7 an 18 $\mu$m are not observed. Iron is used as
the carrier of the warm continuum, but as PAHs have been detected in
\object{NGC~6302} \citep{MLS_01_NGC6302}, amorphous carbon, instead of
iron, could also explain the continuum emission.  A mixed chemistry is
not uncommon for evolved stars, it is for example seen in
\object{IRAS~09425$-$6040} \citep{MYW_01_iras09}. We conclude that the
composition of the warm dust is different from the composition of the
cool dust.

The exact composition of the cold amorphous dust is unknown. Amorphous
olivine is used to produce the required continuum emission, but other
abundant species with a smooth absorption spectrum at far-infrared
wavelengths, such as pyroxene and iron, could be contributing as well.
The mass fractions of forsterite and enstatite are lower limits. We
used the laboratory data leading to the best spectral match with the
observations, i.e.~those of \citet{KTS_99_xsils}. 

\begin{figure*}
  \sidecaption \includegraphics[width=12cm]{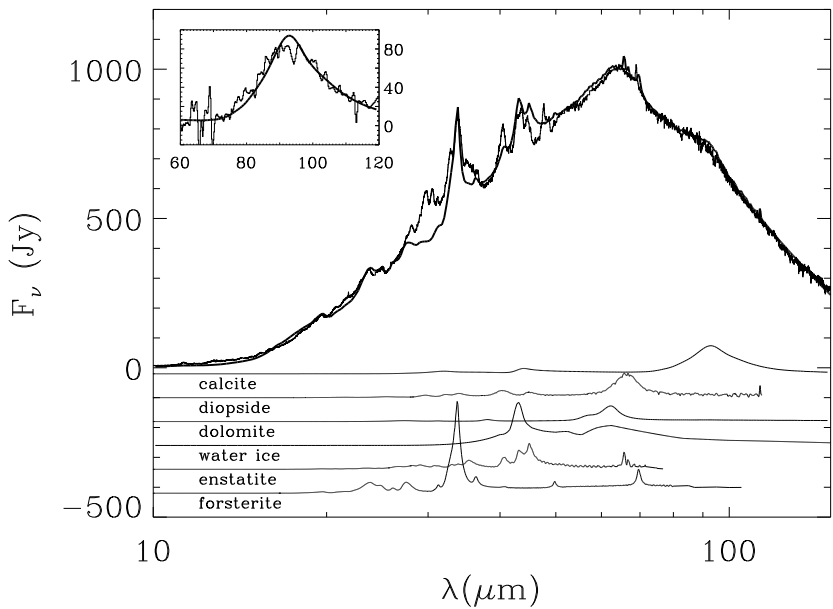}
\caption[]{Model fit to the ISO spectrum of \object{NGC~6302}. The thin line represents
  the observed spectrum, the fit, consisting of the components given
  in Table~\ref{tab:dustmasses}, is indicated with the thick line. In
  the lower part of the plot, the contributions of the cold dust
  components (30--60 K) are indicated, except the contribution of
  amorphous olivine, which leads to a smooth overall contribution. The
  inset shows the observed spectrum (thin line) and the model fit
  (thick line) from 60--120 $\mu$m, where the contributions of all
  species, except calcite, are subtracted from both the observations
  and the model. Thus, it only shows the observed and modelled calcite
  92 $\mu$m feature. Figure adopted from Paper~I.}
\label{fig:fit}
\end{figure*}

The previously unidentified 92 $\mu$m band \citep{MLS_01_NGC6302} can
be fitted very accurately with calcite (CaCO$_3$, trigonal), see
Fig.~\ref{fig:fit}.  The strong resonances of calcite at shorter
wavelengths (see Fig.~\ref{fig:IRtrans}) are not discernible in the
ISO spectrum, indicating that it is not a significant component in the
warm dust. Aragonite (CaCO$_3$, rhombic) also has a resonance at $\sim$92
$\mu$m, and may therefore contribute to the observed feature in
\object{NGC~6302}. However it is not very likely to exist in a
Planetary Nebula.  Although it has the same chemical composition as
calcite, its lattice structure is different, and it can only exist
under relatively high pressure and temperature.

\begin{figure}
  \resizebox{\hsize}{!}{\includegraphics{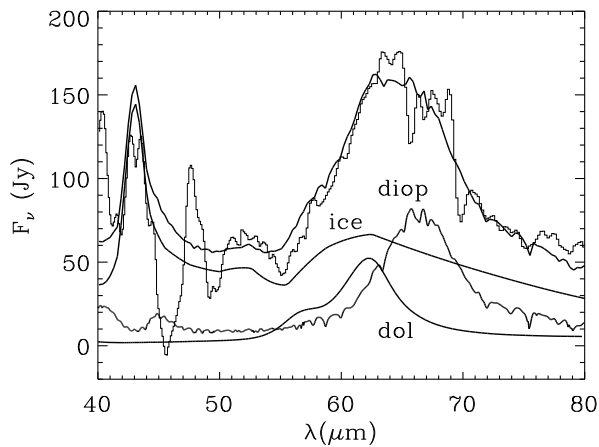}}
\caption[]{Contributions to the 60 $\mu$m complex in \object{NGC~6302}. The thin line
  represents the ISO spectrum of \object{NGC~6302} from 40--80 $\mu$m
  with all contributions subtracted, except the contributions of
  diopside, dolomite and water ice. The thick lines represent the
  contributions of diopside, water ice and dolomite to $F_{\nu}$,
  indicated with abbreviations of their mineral names. The uppermost
  thick line is the sum of the contributions of diopside, dolomite and
  water ice.}
\label{fig:60um}
\end{figure}

The 60 $\mu$m feature can be accurately fitted using absorption
efficiencies of dolomite, diopside and water ice (see
Fig.~\ref{fig:60um}). Ankerite (FeCO$_3$) has a resonance at
approximately the same wavelength (see Fig.~\ref{fig:IRtrans}) and
could in principle contribute to the 60 $\mu$m complex. However, solar
system carbonates are almost exclusively in the form of dolomite and
calcite \citep{JP_93_carbonate,BJ_98_meteorites} and therefore, we
concentrated our analysis on dolomite.

The results of our model fit are shown in Fig.~\ref{fig:fit} and
Table~\ref{tab:dustmasses}.  Since the dust shell is optically thin,
the relative mass fractions can be derived from the model fit.  The
absolute dust mass of each mineralogical component scales with the
distance, and if 910 pc is adopted for the distance \citep{SK_89_PNe},
we arrive at the dust masses as indicated in
Table~\ref{tab:dustmasses}.  We derived that the depletion of calcium
into the identified calcium bearing species -- calcite, dolomite and
diopside -- is about 30\% (Paper~I), if we assume a dust/gas mass
ratio of 1/100 and adopt a calcium abundance of $[$Ca/H$] = 2.2 \cdot
10^{-6}$ \citep{S_84_abundances}.  From the fit to the 60 $\mu$m
complex the fraction of water contained in the solid phase can be
calculated. Adopting a mass ratio of $1 \cdot 10^{-3}$ between H$_2$O
and H$_2$ \citep{GC_99_watervapor} we find that $\sim$ 10\% of the
water is contained in water ice. This is consistent with the results
of \citet{HMD_02_HD161796}. The presence of OH molecules
\citep{PPT_88_NGC6302} indicates that water should also be present in
the vapour phase.

\section{Discussion}
\label{sec:discussion}

\subsection{Geometry and composition of the circumstellar dust shell}
\label{sec:geometry}

In Sect.~\ref{sec:fitparameters} we derived that the dust in the
circumstellar environment of \object{NGC~6302} consists of two
temperature components. The cold component (30--60 K) has a mass of
0.050 $M_{\odot}$ and contains silicates, water ice and carbonates,
indicative of a oxygen-rich chemistry. The warm component (100--118 K)
contains only $1.3 \cdot 10^{-4}$ $M_{\odot}$, which is mainly caused
by a featureless species like carbon or iron. About 5\% of the warm
component is in the form of silicates. The large difference in
chemical composition between the warm and the cold component and the
lack of continuity in the temperature distribution suggest that the
two components have a different spatial distribution. This then allows
a mixed chemistry; regions of C-rich dust could exist in a PN with a
strong silicate spectroscopic signature.  The presence of PAHs in
\object{NGC~6302} \citep[first reported as UIR bands by][]{RA_86_pne}
supports the idea of a mixed chemistry, both the warm, diffuse
component and the PAHs could reside in the same regions of the nebula.

\begin{figure}
\resizebox{\hsize}{!}{\includegraphics{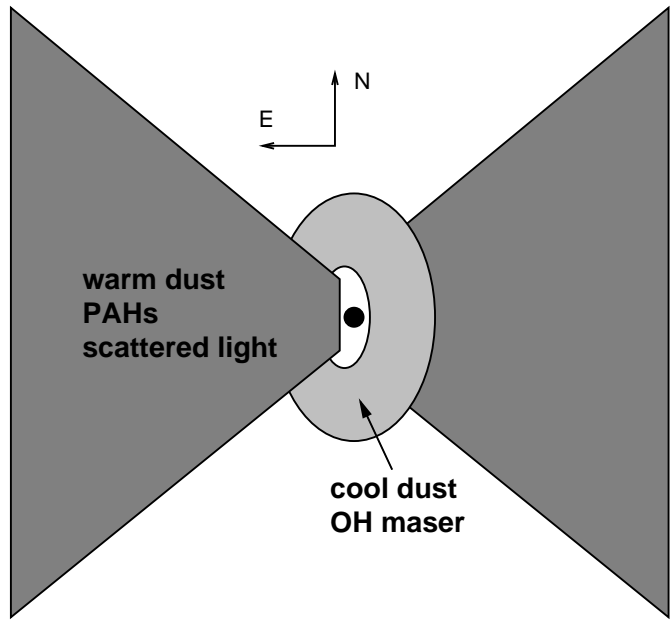}}
\caption[]{Proposed geometry of NGC 6302. A dense torus of cool dust is
  present around the central star (light grey) under an inclination of
  45$^{\mathrm{o}}$ with respect to the line-of-sight. More diffuse
  material in the polar regions, is indicated in dark grey. The torus
  obscures the polar region on the western side. A part of the torus
  is located behind the eastern lobe. The emission from the warm dust
  and PAHs originates from the lobes, as well as scattered light from
  the central star. The torus contains the cool dust and the OH maser.
  For clarity the approximate position of the central star
  is indicated, however in reality it is obscured by the dusty torus
  and can only be studied in scattered light from the polar regions.}
\label{fig:geometry}
\end{figure}

The large mass of the cold component suggests that it is present in a
circumstellar torus, which leaves the polar regions for the warm
carbon-rich component. Optical images of \object{NGC~6302} show that
the eastern lobe is brighter than the western lobe \citep[see
Fig.~\ref{fig:TIMMI2} and also][]{SCM_92_PNimages}, suggesting that
the western lobe is partly obscured by the circumstellar torus.  This
leads to a geometry as indicated in Fig.~\ref{fig:geometry}. The dusty
torus around the central star is seen under an inclination angle $\sim
45^{\mathrm{o}}$, and partially obscures the western lobe. In return,
the eastern part of the torus is located behind the eastern outflow.
The dusty torus contains a high column density of dust towards the
central star, and therefore, most of the dust is effectively shielded
from heating by the central star and remains relatively cool. The dust in
the polar regions however, is directly radiated by the central star
and thus has a higher temperature.

Our observations, as well as previously published work, is in
agreement with the proposed geometry. The torus is optically thick in
the UV, optical and near-infrared. Therefore, the optical images only
show the scattered light from the central star in the polar regions.
We have convolved the warm and cool dust spectra modelled in
Sect.~\ref{sec:fitparameters} with the TIMMI2 N11.9- and Q-band
transmission profile and corrected for the atmospheric transmission in
order to determine the contribution of the dust components in both
bands.  We find that in the N11.9-band, only 0.2\% of the flux comes
from the cold component, and consequently 99.8\% originates from the
warm dust component. For the Q-band these numbers are 29.0\% and
71.0\% respectively.  Since the peak in intensity in the N-band images
and the optical image coincide (see Fig.~\ref{fig:overlay}) we
conclude that indeed the warm dust is located in the polar regions.
The Q-band imaging reveals a different picture; one third of the flux
originates from the cool dust component, and two third from the warm
dust. Both the warm dust in the eastern lobe, and the cool dust in the
western part of the torus contribute to the observed image, leading to
a more or less spherical structure in this wavelength band
(Fig.~\ref{fig:TIMMI2}).

N-band spectroscopy of the eastern lobe shows PAH emission.  From
Fig.~6 of \citep[][figure is rotated by
180$^{\mathrm{o}}$]{CRB_00_ngc6302_ngc6537} it is clear that the
PAH emission intensity peak coincides with the intensity peak in
the optical and N-band, indicating that the PAH emission also
arises from the lobes, where UV radiation from the central star
can penetrate. The UV radiation field also explains the ionization
observed in the lobes (Fig.~\ref{fig:TIMMI2spec}).
\citet{PCM_99_PN} also notice that the emission of warm dust
grains and forbidden line transitions emerges from the same
region. The 6cm data presented by \citet{GMR_89_NGC6302} traces
the location H{\sc ii} region. In their Fig.~3 the region confined
by the torus is clearly visible.

On the other hand, the optically thick torus around the central star
effectively shields the molecular material in the torus from the high
energy radiation. The infrared radiation field in the torus will pump
OH molecules present and cause the formation of OH masers.  Indeed, OH
maser emission is observed in \object{NGC~6302}, on the west side of
the object \citep{PPT_88_NGC6302}, indicating that it arises from part
of the torus located at the front side of the object. This is
consistent with the fact that predominantly the blue-shifted part of
the line profile is seen \citep{PPT_88_NGC6302}.

\subsubsection{Validity of model assumptions}
\label{sec:assumptions}

The geometry described here is consistent with the assumptions we
have used to fit the ISO spectrum, and with the TIMMI2 images.
Here we will discuss the validity of the assumptions used in this
model.

{\bf i)} We assumed that the dust is optically thin at large
wavelengths. The TIMMI2 images indeed suggest that this is true:
the torus is still visible as a dark lane in the N-band images but
starts to disappear when \object{NGC~6302} is observed in the
Q-band. At longer wavelengths, where the spectral energy
distribution (SED) peaks and the different solid state species are
identified, the dust torus will be optically thin.

A simplified calculation verifies this: The dust is thought to be
present in a torus, with a scale height $h$ from the equatorial plane
and an outer radius $R_{\mathrm{out}}$. The inner radius is small
compared to the outer radius and can be ignored.  The torus is seen
under an inclination angle of 45$^{\mathrm{o}}$ \citep{B_94_ngc6302}.
The length of the line-of-sight through the disk is thus $l = 2h
\sqrt{2}$.  Assuming that the density is constant throughout the disk,
we find for the infrared optical depth along this line-of-sight:

\begin{equation}
\tau_{\mathrm{IR}} = Q \,  n \, \sigma_{\mathrm{geom}} \, 2 h \sqrt{2}
\end{equation}

where

\begin{equation}
n = \frac{M_{\mathrm{d}}}{M_{\mathrm{grain}}} \, \frac{1}{2h \pi {R_{\mathrm{out}}}^2}
\end{equation}

By adding all cold dust components listed in
Table~\ref{tab:dustmasses}, the dust contained in the torus is
determined to be $M_{\mathrm{d}} = 0.05$ $M_{\odot}$.  The mass of a
grain follows from the used grain size of 0.1 $\mu$m, and the average
density of silicate grains. Using cgs units, we find that the torus is
optically thin in the mid- and far-infrared in case

\begin{equation}
\frac{Q_{\mathrm{IR}}}{{R_{\mathrm{out}}}^2} < 1.1 \cdot 10^{-36} \, \mathrm{cm}^{-2}
\end{equation}

For a typical value of $Q_{\mathrm{IR}}$ in the mid- and far-infrared
of 0.0028 we find that for $R_{\mathrm{out}} > 5.0 \cdot 10^{16}$ cm,
or $3.4 \cdot 10^3$ AU, the dusty torus is optically thin at these
wavelengths.  This size is not unreasonable for a circumstellar
torus, as it is comparable to the sizes of the dust shell shells
around AGB stars, the PN progenitors. In addition,
we can argue that the density distribution will not be flat, but
probably behaves more like a power law.  Then, most lines-of-sight
through the torus will be optically thin even in case of a smaller
torus. Therefore, we are confident that the torus is indeed optically
thin in the mid- and far-infrared.

{\bf ii)} We assumed that all grains have a size of 0.1 $\mu$m. In
reality, the grains formed in the circumstellar environment will cover
a wide range of sizes. However, we believe that we actually get a good
estimate of the actual dust masses, because we use the mass absorption
coefficients, and is thus independent from grain size. Using the
mass absorption coefficients will lead to a direct determination of
the dust mass, without calculating the mass of individual grains. The
calcite and dolomite data presented in this work are mass absorption 
coefficients $\kappa_\nu$, the data
used for diopside, enstatite and forsterite are measured in terms of
absorption efficiency $Q_\nu/a$.  For amorphous olivine, metallic iron and
water ice we used optical constants, and derived absorption efficiency
for grains with a size of 0.1 $\mu$m. Only the mass determination of
these three species is probably somewhat affected by our assumption on
the grain size.

{\bf iii)} We assumed that all grains are spherical. Again an
assumption that is probably not very realistic. The dominant dust
component (amorphous olivine) must be in the form of non-spherical
particles to explain the infrared-spectrum of AGB stars, and the same
is true for metallic iron \citep{KDW_02_composition}. The similarity
in the spectral appearance of the solid state features detected in PNe
and their direct progenitors, AGB stars,
\citep[e.g.][]{SKB_99_ohir,MWT_02_xsilI}, indicates that the dust
grains in the two types of objects probably have similar properties.
Therefore, we may conclude that the grains in \object{NGC~6302} are
very likely not homogeneous spheres, as these particles result in a
shift in peak position with respect to non-spherical particles and
inhomogeneous spheres.  However, in our model the only purpose of the
grain shape is to calculate the total mass of some of the dust
components. The effect that grain shape has on light scattering is not
taken into account, as our model uses mass absorption coefficients.

{\bf iv)} We assumed that all grains are of homogeneous composition.
Using mass absorption coefficients to calculate the emerging spectrum,
we simply derived the total mass of a certain mineral found in the
line of sight. It is impossible to discriminate between homogeneous
grains, and grains consisting of two or more mineral species,
therefore we cannot exclude the presence of grains of mixed
composition. For instance, it is well known that volatiles, such as
water vapour, condense in layered structures on grains
\citep{JM_85_condensation}. We stress that the derived total masses
are independent from the assumed homogeneity of grains.

{\bf v)} We assumed that the density and temperature gradients are
simple power laws of the distance to the central star.  The complex
structure of \object{NGC~6302} indicates that this is probably far
besides the truth, but since the optical depth is low, all infrared
radiation emitted by the grains is received. Therefore not the density
and temperature as a function of distance to the central star, but the
mass-temperature relation of the different dust species becomes
important, in a similar fashion as described by \citet{BDV_00_ABAur}.
Our simulation of this relation by a power law density and temperature
gradient seems to be in agreement with the observed spectrum.

{\bf vi)} Finally we assumed that all dust species are found in the
same temperature range. However, if the different dust components are
not in thermal contact with each other, i.e.~if they are present in
separate -- but co-spatial -- grain populations, the difference in
optical properties, notably in the UV and visual, causes a difference
in the temperature profile \citep{BDV_00_ABAur,KWD_01_xsilvsmdot}. The
determination of the temperature range is based on the overall shape
of the SED, and we forced the crystalline components -- which are the
carriers of the narrow features superposed on the broad continuum --
to have the same temperature profile as the amorphous component. Since
the dust emission strongly depends on the temperature of the grains,
this will probably be our main source of error in our mass estimates.
However, dropping the constraint that all dust species are found in
the same temperature range, leads to a large increase in free
parameters for our simple model.

Considering all the effects discussed here, we estimate that our
absolute mass determinations are accurate within a factor of two, if
the distance toward \object{NGC~6302} is 910 pc. If we take into
account the large uncertainty in the distance determinations, the
error in the absolute dust mass determination further increases.  The
uncertainty in the various laboratory measurements introduces an
additional uncertainty of a factor of two, leading to a total error of
a factor of $\sim$3.  This inaccuracy in the laboratory measurements
can be inferred from comparison between published data.  The
crystalline silicate features published by \citet{KTS_99_xsils} are
intrinsically much stronger (up to 5 times), than the same features in
measurements by other groups
\citep{SP_73_mg2sio4,S_74_optprop,JMD_98_crystalline}.  Consequently,
the crystallinity, defined by the mass of enstatite and forsterite as
a fraction of the total dust mass, lies within 5\% -- 25\%, and the
values given in Table~\ref{tab:dustmasses} are thus just lower limits.
The error in the relative dust mass fractions, i.e.~the mass ratios
between the identified minerals, is much better, as they all depend in
the same way on our model assumptions. We therefore estimate the error
in the relative dust mass fractions of the order of 50\%.  The error
in the absolute and relative flux levels is small compared to the
inaccuracies discussed here, and is therefore not taken into account.

In order to further approach reality, full radiative transfer
calculations should be performed.  Using spatial information derived
from imaging and measurements of outflow velocities it is possible to
reconstruct the geometry of the nebula. With 2-dimensional full
radiative transfer calculations over a large wavelength range, from UV
to far-infrared, it is possible to resolve the temperature and density
distribution throughout the PN, for the different dust components.
Also, the effect of grain sizes, grain shapes and mineralogical
homogeneity then becomes apparent in the spectroscopy. This will lead
to a much more accurate description of the properties of the dust in
the nebula. However, we chose not to do these calculations for the
following reasons: First, optical constants are not available for some
of the identified dust components. Second, our knowledge about the
geometry of the nebula is still too limited to justify the huge effort
of 2-dimensional radiative transfer calculations. These calculations
introduce new free parameters, and it will be very hard to constrain
the model parameters in an unambiguous way with the limited
information available. Spatially resolved spectroscopy of the mid- and
far-infrared regions will certainly improve the image we have of
\object{NGC~6302}. Finally, to obtain reliable mass estimates the
distance toward \object{NGC~6302} should be more accurately known.

\subsection{The astronomical relevance of carbonates}
\label{sec:literature}

In paper~I and this work, we present the first extrasolar
detection of carbonates. The detection is based on far-infrared
features of carbonate at 92 $\mu$m and dolomite, which contributes
to the 60 $\mu$m complex. Due to the low abundances, the strong
resonances present at short wavelengths (see
Fig.~\ref{fig:IRtrans}) do not stand out in the spectra, as many
other species contribute significantly to the opacity at these
wavelengths.  This is not the case at the far-infrared
wavelengths, therefore it is possible to detect even very small
amounts of carbonates, beyond reasonable doubt.

There have been a number of searches for carbonate features in the
mid-infrared. The previously unidentified infrared (UIR) feature
at 11.3 $\mu$m seen in planetary nebula \object{NGC~7027} was
attributed to carbonates \citep{GFM_73_PNe,BR_75_carbonate}.
However, this identification became unlikely with the
non-detection of the 6.8 $\mu$m carbonate feature in
\object{NGC~7027} \citep{RSW_77_NGC7027}, and the lack of
characteristic features due to carbonates in the 22--35 $\mu$m
region of the spectrum \citep{MFH_78_NGC7027}.  The UIR feature at
11.3 $\mu$m in \object{HD~44179}, also known as the Red Rectangle,
was attributed to the carbonate magnesite (MgCO$_3$)
\citep{B_77_HD44179}.  However, \citet{CAT_86_COratio} showed PAHs
are the most likely carrier of the UIR feature at 11.3 $\mu$m seen
in various astrophysical environments.  Thus, the sole detection
of the 11.3 $\mu$m feature is no longer considered evidence for
the presence of carbonates.

Interesting environments to search for carbonates are star forming
regions and young stars. Carbonates form easily on the surface of
planets when liquid water is present, but other formation mechanisms
cannot be excluded (Paper~I). \citet{RSP_77_BNKL} have determined an
upper limit to the carbonate/silicate mass ratio in the interstellar
medium toward Orion. They find that the this mass ratio is at most
0.05, based on the detection limit of the 6.8 $\mu$m feature. The 6.8
$\mu$m feature is seen in absorption toward embedded protostars. The
resemblance of the spectral appearance of the 6.8 $\mu$m feature in
\object{W33A} and in Interplanetary Dust Particles (IDPs) has lead to
the conclusion that they have the same carrier, most likely carbonates
\citep{SW_85_laboratory,TB_86_IDP}, although \citet{TAB_84_protostars}
argue that it is due to hydrocarbons.  \citet{C_89_AFGL961} observes
the 6.8 $\mu$m feature towards protostar \object{AFGL~961} and claims
that it is partially due to carbonates along with a contribution of
hydrocarbons.  Hints of the 11.3 and 13.4 $\mu$m carbonate features
are also claimed to be present.  The similarity between the 6.8 $\mu$m
features in IDPs and in interstellar lines-of-sight is reported by
\citet{QRB_00_7micron} as well, presumably coming from the same
carrier, although carbonates are rejected as a carrier.  From the high
resolution ISO SWS spectroscopy it becomes clear that the spectral
shape of the 6.8 $\mu$m absorption feature observed towards protostars
cannot be explained by carbonates, but a satisfactory alternative is
still lacking \citep{KTB_01_ice}. Recently, the $\sim$92 $\mu$m
carbonate feature has been detected towards protostar
\object{NGC~1333-IRAS~4} (Ceccarelli et al., in prep.).

Although extraterrestrial carbonates are quite often found in
meteorites and IDPs, they are actually not easy to detect by means of
astronomical observations.  The only claim still standing at the
moment is the detection of features at 26.5 and 31 $\mu$m due to
dolomite and calcite respectively in the ISO SWS spectrum of
\object{Mars} \citep{LED_00_Mars}. However, the far-infrared features
of dolomite and calcite at $\sim$60 and $\sim$92 $\mu$m are not seen
on \object{Mars} \citep{BEB_00_marsFIR}.

As pointed out in Paper~I, until now carbonates are believed to be
formed through aqueous alteration. In this formation process, carbon
dioxide (CO$_2$) dissolves in liquid water and forms carbonate ions
(CO$_3$$^{2-}$). If cations like Ca$^{2+}$ or Mg$^{2+}$ are found in
the solution as well, carbonates can be formed as a lake sediment when
saturation is reached. The presence of carbonates is seen as evidence
of planet formation, as an atmosphere containing carbon dioxide is
required, as well as liquid water on the surface of the planet. The
connection with planet formation is the justification for the
 search for carbonates in the circumstellar environment of
young stars. 
Carbonates are ubiquitous in our own solar system, as is
shown from the composition studies of meteorites and interplanetary
dust particles. Carbonate-containing meteorites are considered tracers
of the formation history of the solar system
\citep[e.g.][]{EZB_96_aqueous,B_98_aqueous}. However, the detection of
carbonates in PNe suggests that alternative formation mechanisms exist
and that the presence of carbonates no longer provides direct evidence
for planet formation (Paper~I).

\section{Summary}
\label{sec:summary}

We have determined the composition and distribution of the dust in the
planetary nebula \object{NGC~6302}. We found that a warm (100-118 K)
and a cool (30-60 K) dust component are present. The cool dust
component is located in a circumstellar torus of which the inner part
effectively shields the UV/optical and near-infrared radition from the
central star.  The torus contains amorphous olivine, forsterite,
clino-enstatite, water ice, diopside, dolomite and calcite.  Outside
the solar system, the carbonates dolomite and calcite are only seen in
\object{NGC~6302} and \object{NGC~6537} (see also Paper I), and the
formation mechanism of carbonates in these environments is not yet
understood.  Diopside is so far only found in \object{NGC~6302}
\citep{KTS_00_diopside} and in two OH/IR stars \citep{DDW_00_OHIR}.
Water ice, forsterite, enstatite and amorphous olivine are very common
in the dust shells of evolved stars
\citep[e.g.][]{MWT_02_xsilII,MWT_02_xsilI,MWT_02_xsilIII}.  The
circumstellar torus also contains the OH maser reported by
\citet{PPT_88_NGC6302}.

The optical depth in the polar regions is smaller, and therefore, these
regions can be observed in the optical in
scattered light from the central star. These regions contain the warm
dust component, which thermally emits in the mid-infrared. The N-band
images are dominated by thermal emission from the warm dust. From the
ISO spectroscopy, constraints on the dust composition could be derived:
predominantly metallic iron (or another featureless dust component), amorphous
olivine, forsterite
and clino-enstatite are detected. For water ice, diopside, calcite and
dolomite we have derived upper limits for the mass fraction in the
warm component. Only 0.3\% of the total dust mass is contained in the
warm component, which has a different mineralogical composition than the
cool component. UV radiation can penetrate the polar regions and
thus we are able to explain the observed PAH emission, which is
also reported by \citet{CRB_00_ngc6302_ngc6537}.

In addition, the mass absorption coefficients of carbonate minerals are
presented in this work. We studied calcite, dolomite, ankerite, aragonite
siderite and magnesite.

\acknowledgements{We thank M.~Matsuura for her help in obtaining the
  TIMMI2 observations and A.~de Koter for the useful discussions. We are
  grateful for the support by the staff of the ESO 3.6m telescope.
  LBFMW and FK acknowledge financial support from an NWO 'Pionier'
  grant.  This work was partly supported by NWO Spinoza grant 08-0 to
  E.P.J.  van den Heuvel.}

\bibliographystyle{apj}
\bibliography{ciska}

\end{document}